# Seamless Integration and Implementation of Distributed Contact and Contactless Vital Sign Monitoring

Dingding Liang[1,†], Yang Chen[1,†,*], Jiawei Gao[1], Taixia Shi[1], and Jianping Yao[2,*]

[1] Shanghai Key Laboratory of Multidimensional Information Processing, School of Communication and Electronic Engineering, East China Normal University, Shanghai 200241, China

[2] Microwave Photonic Research Laboratory, School of Electrical Engineering and Computer Science, University of Ottawa, Ottawa, ON K1N 6N5, Canada

[†]These authors contributed equally to this work.

[*] Correspondence to: Y. Chen, ychen@ce.ecnu.edu.cn; J. Yao, jpyao@uottawa.ca.

**ABSTRACT**
Real-time vital sign monitoring is gaining immense significance not only in the medical field but also in personal health management. Facing the needs of different application scenarios of the smart and healthy city in the future, the low-cost, large-scale, scalable, and distributed vital sign monitoring system is of great significance. In this work, a seamlessly integrated contact and contactless vital sign monitoring system, which can simultaneously implement respiration and heartbeat monitoring, is proposed. In contact vital sign monitoring, the chest wall movement due to respiration and heartbeat is translated into changes in the optical output intensity of a fiber Bragg grating (FBG). The FBG is also an important part of radar signal generation for contactless vital sign monitoring, in which the chest wall movement is translated into phase changes of the radar de-chirped signal. By analyzing the intensity of the FBG output and phase of the radar de-chirped signal, real-time respiration and heartbeat monitoring are realized. In addition, due to the distributed structure of the system and its good integration with the wavelength-division multiplexing optical network, it can be massively scaled by employing more wavelengths. A proof-of-concept experiment is carried out. Contact and contactless respiration and heartbeat monitoring of three people are simultaneously realized. During a monitoring time of 60 s, the maximum absolute measurement errors of respiration and heartbeat rates are 1.6 respirations per minute and 2.3 beats per minute, respectively. The measurement error does not have an obvious change even when the monitoring time is decreased to 5 s.

## 1. Introduction

Vital signs, including respiration rate, heartbeat rate, blood pressure, body temperature, etc., are a crucial set of indicators for evaluating one's physical state[1]. Abnormal respiration and heartbeat rates are often early indicators of various illnesses or health problems such as lung disease, cardiovascular disorders, and epileptic seizures[2–4]. Real-time monitoring of respiration and heartbeat can facilitate early intervention, help individuals adjust their lifestyle, and improve diet and exercise habits, which are important for the health of modern humans and are receiving increasing attention. Respiration and heartbeat monitoring typically fall into two categories: contact and contactless. Conventional contact respiration or heartbeat monitoring devices, including the electrocardiograph (ECG)[5], the photoplethysmography (PPG)[6], the stethoscope, Chinese medicine pulse diagnosis, and the smartwatch[7], face several issues. For instance, ECG and PPG require long-term placement of probes on the skin, the stethoscope and Chinese medicine pulse diagnosis methods are extremely labor-intensive, and the accuracy of heartbeat rate measurement by smartwatches needs to be improved. Camera-based and radar-based monitoring are two commonly employed contactless methods for monitoring respiration and heartbeat[8–11]. Camera-based monitoring is not only sensitive to lighting conditions but also raises privacy concerns[12]. Radar-based monitoring transmits large-bandwidth radar signals to achieve high-precision range measurement and then uses various algorithms to extract respiration and heartbeat rates[13–16]. Since the accuracy of radar-based monitoring is closely related to the radar signal bandwidth, to achieve high accuracy, it is necessary to use a large-bandwidth radar signal. Regrettably, broadband radar signal generation and processing are inherently limited by electronic bottlenecks[17].

To address the limitations of conventional contact monitoring devices, a range of optical fiber sensing systems have been proposed for physical state monitoring, taking advantage of their favorable attributes such as compact size, lightweight design, and high sensitivity[3,18–20]. A wearable optical fiber sensing system combines fiber Bragg grating (FBG) and a fiber laser sensor to measure multiple physiological parameters of the human body[21]. The optical fiber sensing system has a simple structure and can monitor the physical state in real-time. In ref. [22], different respiration patterns such as exhalation and inhalation are identified accurately by a tiled FBG, whose grating planes are inclined relative to the axis of the fiber. In ref. [23], a helical long-period fiber grating sensor system is established for simultaneously monitoring respiration and heartbeat rates. The maximum absolute measurement errors of respiration and heartbeat rates of four healthy subjects are 2 respirations per minute (rpm) and 3 beats per minute (bpm), respectively. In ref. [24], an FBG-based wearable sensor is constructed with a high sensitivity to simultaneously monitor respiration and heartbeat activities. The respiration and heartbeat rates of five participants are obtained in different positions, such as standing, sitting, and supine. Compared with conventional contact respiration or heartbeat monitoring devices, fiber-grating-based contact respiration or heartbeat monitoring systems have simple structures and good monitoring accuracy, can reduce labor consumption, and are more convenient and comfortable for long-term monitoring[21–24].

In contactless respiration and heartbeat monitoring, radar-based monitoring not only achieves all-time human physiological parameters monitoring but also has stronger privacy protection than the camera-based vital sign monitoring approach[25]. To escape from the electronic bottlenecks, microwave photonic radar has been widely studied to generate and receive broadband radar signals[26–29]. In the transmitter, high-frequency and broadband radar signals are generated by photonic frequency multiplication[30,31], photonic frequency conversion[32], optical injection[33,34], photonic digital-to-analog converter[35], or acoustic-optic frequency shifting loops[36,37]. In the receiver, the photonic de-chirping process is employed to convert high-frequency and broadband linearly frequency-modulated (LFM) signals into low-frequency and narrowband electrical signals[26,32]. At present, the research of microwave photonic radar is mainly focused on high-accuracy target ranging and imaging, and its application to vital sign monitoring is relatively rare. In 2023, a microwave photonic radar for contactless vital sign monitoring is proposed to monitor the respiration of the simulators and a cane toad[37]. The synthesized stepped frequency signal with a bandwidth of 10 GHz is generated by using an acoustic-optical frequency shifter, and the photonic de-chirping process is utilized to extract the respiration rate. The chest wall movements of the two breathing simulators are successfully extracted, and the corresponding respiration rates are 12 rpm and 16.5 rpm, respectively. The cross-correlation coefficient between the cane toad's radar and camera measurements is 0.746.

In medical monitoring, contact respiration and heartbeat monitoring are more popular, but in some cases, contactless vital sign monitoring may be more suitable for patients who require long-term monitoring such as premature infants and burn patients[38]. In addition, contactless vital sign monitoring is more convenient and faster for some people who perform routine inspections[39]. Therefore, both contact and contactless vital sign monitoring play an important role. With the steady growth of population aging and the increasing awareness of physical health, large hospitals, nursing homes, and smart communities all urgently need vital sign monitoring systems to detect the vital signs of patients, the elderly, and everyone in the community, so as to quickly and accurately obtain vital sign information for different types of patients, help continuously monitor the health state of the elderly and everyone in the community, and maintain their physical health[40]. Typically, contact and contactless vital sign monitoring systems are developed independently, with limited research exploring their potential for integration. However, regardless of whether it is contact, contactless, or integrated, installing a set of vital sign monitoring equipment in every consulting room and ward of the hospital, every room of the nursing home, or every family in the community is very expensive. In response to the needs of future smart and healthy cities, there is a strong desire for a technology that can integrate contact and contactless vital sign monitoring, and achieve low-cost, large-scale deployment, thereby meeting the requirements of different application scenarios, user groups, and deployment locations in the future.

The microwave photonic technology used to realize the above-mentioned microwave photonic radar is a technique that combines microwave engineering and photonic technology to take advantage of the wide bandwidth offered by photonics for microwave signal generation and processing[41–43]. It is possible to fulfill the needs by combining microwave photonic technology with the existing sophisticated wavelength-division multiplexing (WDM) optical network infrastructure and radio-over-fiber (ROF) technique[44]. In this article, for the first time, we propose a seamlessly integrated contact and contactless vital sign monitoring system that utilizes microwave photonics and achieves the distribution of optical vital sign monitoring signals in a WDM optical network through ROF technology, enabling simultaneous distributed respiration and heartbeat monitoring. Figure 1 shows the application scenario of the proposed vital sign monitoring system in future smart and healthy cities. Large hospitals, nursing homes, smart communities, and the central office are connected through a WDM optical network. Each user terminal of hospitals, nursing homes, and smart communities has been equipped with contact and contactless vital sign monitoring functions. The optical vital sign monitoring signal distributed in the WDM optical network is used for both contact and contactless respiration and heartbeat monitoring. In contact respiration and heartbeat monitoring, the chest wall movement is translated into changes in the optical intensity of the carrier of the optical vital sign monitoring signal, via an FBG. The optical vital sign monitoring signal output by the FBG is subjected to optical filtering and an optical-to-electrical converter to generate a frequency-quadrupled LFM signal as the broadband radar signal for contactless monitoring, in which the chest wall movement is translated into phase changes of the radar de-chirped signal. By analyzing the intensity of the FBG output and phase of the radar de-chirped signal, contact and contactless respiration and heartbeat information can be obtained simultaneously. Due to the complex and expensive equipment, such as light sources and wideband radio frequency sources, being centralized at the central office, the cost for each user accessing via the WDM optical network is significantly reduced, and the number of users supported by the proposed system can be greatly expanded by increasing the number of optical wavelengths. A proof-of-concept experiment shows that contact and contactless vital sign monitoring of three people are simultaneously realized. During a monitoring time of 60 s, the maximum absolute measurement errors of respiration and heartbeat rates are 1.6 rpm and 2.3 bpm, respectively. The measurement error does not have an obvious change even when the monitoring time is decreased to 5 s.

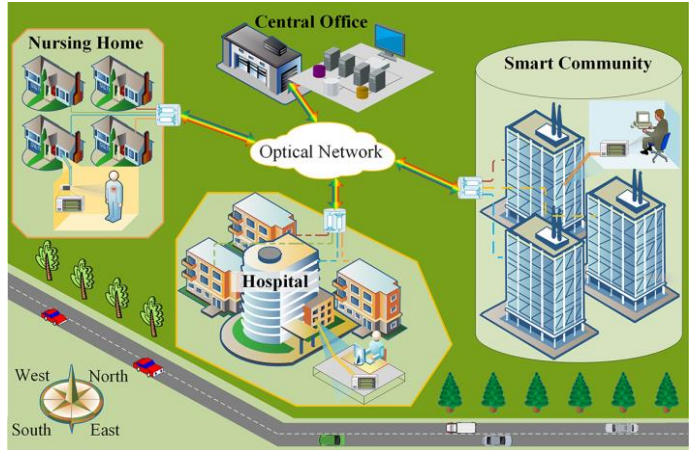

**Fig. 1 Application scenario of the vital sign monitoring in future smart and healthy cities.**

## 2. Principle and system

Figure 2 shows the schematic diagram of the proposed seamlessly integrated contact and contactless vital sign monitoring system with ROF assistance incorporated into a WDM optical network. The vital sign monitoring system mainly consists of a central office and multiple user terminals connected by the WDM optical network. Each user terminal is equipped with one or both functions of FBG-based contact respiration and heartbeat monitoring and radar-based contactless respiration and heartbeat monitoring. Different respiration and heartbeat monitoring methods can be selected according to the application scenario and the user state.

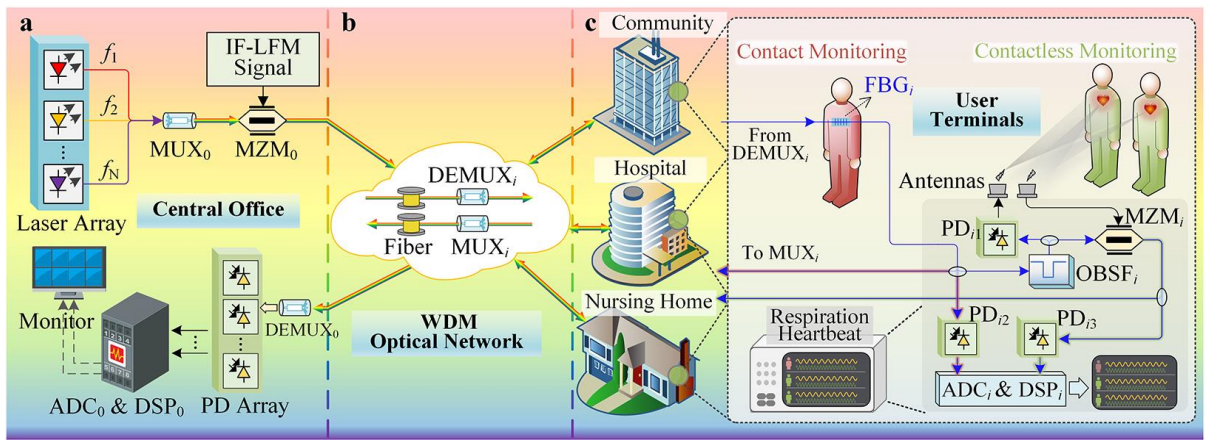

**Fig. 2 Schematic diagram of the proposed seamlessly integrated contact and contactless vital sign monitoring system with ROF assistance incorporated in a WDM optical network. a** Central office. **b** WDM optical network. **c** User terminals. MUX optical multiplexer, MZM Mach–Zehnder modulator, DEMUX optical demultiplexer, FBG fiber Bragg grating, OBSF optical band-stop filter, PD photodetector, ADC analog-to-digital converter, DSP digital signal processing.

In the central office, as shown in Fig. 2a, multiple optical carriers ($f_1$, $f_2$, ···, $f_N$), in accordance with ITU WDM wavelength standards, are generated from a laser array and then injected into a Mach–Zehnder modulator (MZM$_0$) after wavelength multiplexing in an optical multiplexer (MUX$_0$). In MZM$_0$, each optical carrier is modulated independently by an intermediate frequency (IF)-LFM signal. The instantaneous frequency of the IF-LFM signal is denoted as $f_{IF}(t) = f_C - 0.5f_B + kt$, where $f_C$, $f_B$, and $k$ are the center frequency, bandwidth, and chirp rate of the IF-LFM signal, respectively. MZM$_0$ is biased at the maximum transmission point (MATP) so that each optical carrier is modulated to produce only the even-order optical sidebands of the IF-LFM signal. Under the small-signal modulation condition, only the original optical carriers and their ±2nd-order optical sidebands are taken into consideration, as displayed in Fig. 3a. The multiple optical carriers after modulation in MZM$_0$ are transmitted in an optical fiber to different operational locations, such as hospitals, nursing homes, and smart communities. In each operational location, an optical demultiplexer (DEMUX) is used to separate the different WDM channels, as shown in Fig. 2b. Based on the specific requirements of each user terminal for respiration and heartbeat monitoring, one or more wavelengths are transmitted to the user terminal through short optical fibers. The user terminal can be a consulting room or ward in a hospital, a bedroom or activity room in a nursing home, or a house or apartment in a smart community. After this short-distance fiber distribution, at least one optical carrier centered at $f_i$ and its two ±2nd-order optical sidebands $f_i \pm (2f_C - f_B + 2kt)$ are deployed in one user terminal.

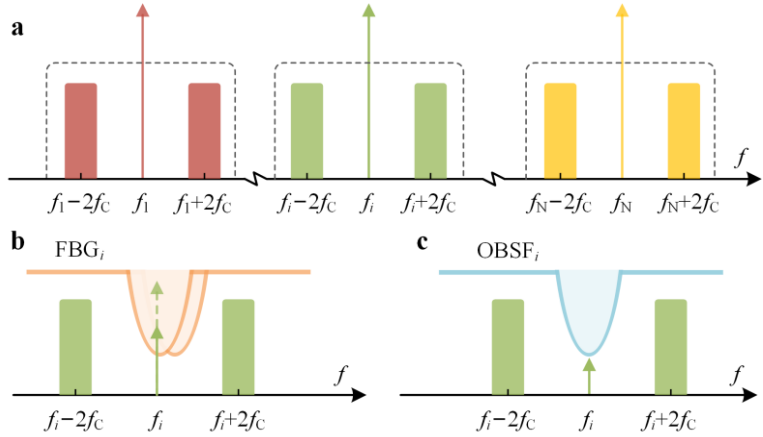

**Fig. 3 Diagrams of key nodes in the proposed vital sign monitoring system. a** Optical spectrum of the IF-LFM-modulated signal from MZM$_0$. **b** Optical spectrum of FBG$_i$ output signal. **c** Optical spectrum of OBSF$_i$ output signal.

In a user terminal, as shown in Fig. 2c, the assigned optical wavelength from DEMUX$_i$, as well as its two optical sidebands, is connected to FBG$_i$, which is used as a contact sensor and taped to the human chest to measure respiration and heartbeat. To implement contact monitoring, the optical carrier is placed at a falling edge of FBG$_i$ transmission notch, as illustrated in Fig. 3b. The chest wall movement caused by respiration and heartbeat introduces the Bragg wavelength shift of FBG$_i$. Note that FBG$_i$ used here should have a narrow transmission notch, which should be no more than $2f_C$. In this way, only the optical carrier is within the transmission notch, while the two 2nd-order optical sidebands of the IF-LFM signal are outside the transmission notch. Under these circumstances, the Bragg wavelength shift introduced by the chest wall movement will directly affect the intensity of the optical signal passing through FBG$_i$. Then, one part of the optical signal passing through FBG$_i$ is converted into an electrical signal in a low-speed photodetector (PD$_{i2}$) to monitor the optical signal intensity that is changed with respiration and heartbeat. By processing the electrical signal from PD$_{i2}$, one's respiration and heartbeat rates can be obtained. At the same time, another part of the optical signal can be transmitted back to the central office via the WDM optical network, so that the optical signal carrying the respiration and heartbeat information from different user terminals can also be processed and monitored centrally.

The last part of the optical signal from FBG$_i$ is sent to an optical band-stop filter (OBSF$_i$) to suppress the optical carrier, as shown in Fig. 3c. To reduce the cost, the OBSF$_i$ can also be a narrowband FBG. The optical signal from the OBSF$_i$ with carrier suppressed is split into two parts: One part is beaten in a high-speed PD$_{i1}$ to generate a frequency-quadrupled LFM signal; The other part is used as an optical reference for radar de-chirping. The instantaneous frequency of the generated frequency-quadrupled LFM signal is expressed as $f_T(t) = 4f_C - 2f_B + 4kt$. The LFM signal is sent to the free space through an antenna to achieve contactless vital sign monitoring. The echo signal reflected by the human chest is collected by the receiving antenna with a time delay $\Delta\tau = 2(R_0 + x(t))/c$, where $R_0$ is the range between the radar antenna and the center of the human chest, $x(t)$ is the chest wall movement produced by respiration and heartbeat, and $c$ is the speed of light. The echo signal is sent to MZM$_i$, which is biased at the quadrature transmission point (QTP) and modulates the optical reference from the OBSF$_i$. The optical signal after modulation is beaten in another low-speed PD$_{i3}$ to generate the radar de-chirped signal in the user terminal or transmitted to the center station after fiber distribution. The de-chirped signal can be expressed as $E_{\text{IF}}(t) \approx E_0\cos(8\pi k\cdot\Delta\tau\cdot t + 4\pi(R_0 + x(t))/\lambda_c)$, where $\lambda_c = c/(4f_C - 2f_B)$, $E_0$ is the amplitude of the de-chirped signal. The frequency and phase of the de-chirped signal are denoted as $f_{\text{IF}} \approx 4k\Delta\tau$ and $\varphi_{\text{IF}} = 4\pi(R_0 + x(t))/\lambda_c$. As can be seen, the chest wall movement of the human body is converted into phase changes of the radar de-chirped signal that can be written as $\Delta\varphi_{\text{IF}} = 4\pi\cdot\Delta x(t)/\lambda_c$. Therefore, the human's position can be determined by the frequency of the de-chirped signal, and the respiration and heartbeat information can be extracted from the phase of the radar de-chirped signal.

## 3. Experimental setup and results

### 3.1 Experimental setup

To demonstrate the proposed seamlessly integrated contact and contactless vital sign monitoring system, a proof-of-concept experiment is carried out based on the setup shown in Fig. 4. Two wavelengths are employed in the experiment: One for contact respiration and heartbeat monitoring and the other for simultaneous contact and contactless respiration and heartbeat monitoring. Two continuous-wave (CW) laser sources (ID Photonics, CoBriteDX1-1-C-H01-FA and CoBriteDX1-1-HC1-FA) centered at 1549.36 nm and 1549.92 nm, respectively, are employed. Due to the accuracy of FBG fabrication, in the proof-of-concept experiment, the optical wavelengths we selected do not strictly conform to the optical wavelength and wavelength spacing of ITU WDM standard, but it will not affect the verification of the method and concept in Fig. 2 through this experiment. The two optical wavelengths are combined in an optical coupler (OC$_1$) after passing through two polarization controllers (PC$_1$ and PC$_2$), which are utilized to adjust their polarization states. Here, the function of OC$_1$ is similar to that of the MUX in Fig. 2. Since there is a lack of suitable MUX in the laboratory, OC$_1$ is used as a substitute. The combined optical signal from OC$_1$ is injected into MZM$_0$ (Fujitsu FTM7938EZ). MZM$_0$ is driven by an IF-LFM signal with a 6.6-GHz center frequency and a 1-GHz bandwidth. The IF-LFM signal is generated from a microwave up-conversion module, which includes an arbitrary waveform generator (AWG, Keysight M8190A), a microwave signal source (MSG, Agilent 83752B), a mixer (MITEQ M30), EA$_1$ (CTT ALM/145-5023-293), and EBPF$_1$ (KGL YA353-2). The optical signal after modulation in MZM$_0$ is transmitted through a 4.1-km optical fiber and then divided into two parts by OC$_2$. Here, the function of OC$_2$ is expected to be similar to that of the DEMUX in Fig. 2. Since OC$_2$ does not have the filtering and wavelength selection functions of the DEMUX, when the optical signal from it is used to generate radar signals for subsequent contactless respiration and heartbeat monitoring, an additional optical filter is required to select the corresponding wavelength. Nevertheless, for contact respiration and heartbeat monitoring, filtering is not necessary because FBG only operates on its corresponding wavelength.

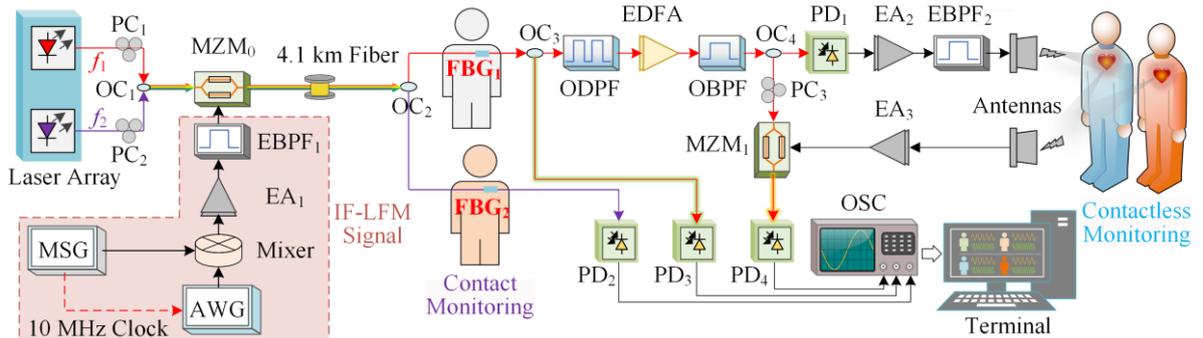

**Fig. 4 Experimental setup of the contact and contactless vital sign monitoring system.** PC polarization controller, OC optical coupler, MZM Mach–Zehnder modulator, EBPF electrical band-pass filter, EA electrical amplifier, MSG microwave signal generator, AWG arbitrary waveform generator, FBG fiber Bragg grating, ODPF optical dual-passband filter, EDFA erbium-doped fiber amplifier, OBPF optical band-pass filter, PD photodetector, OSC oscilloscope.

The two outputs of OC$_2$ are sent to FBG$_1$ and FBG$_2$, which are two contact sensors and attached to the human chest for contact respiration and heartbeat monitoring. The transmission outputs of the two FBGs are detected in two low-speed PDs (PD$_2$ and PD$_3$, Beijing Lightsensing Technologies Ltd. BLPD-PFA1-60BR-W7) to convert the optical power change to the electrical domain. By analyzing the power of the FBG output, the respiration and heartbeat monitoring of two people can be achieved at the same time. The output of $FBG_1$ is divided into two parts by $OC_3$: One part is used for the aforementioned contact respiration and heartbeat monitoring; The other part is used for subsequent contactless respiration and heartbeat monitoring.

The optical signal for contactless respiration and heartbeat monitoring from $OC_3$ is connected to an optical dual-passband filter (ODPF, Finisar WaveShaper 4000A). The ODPF not only functions as the DEMUX in Fig. 2 to select the corresponding WDM channel but also serves as the OBSF in Fig. 2 to suppress the optical carrier. Therefore, the ±2nd-order optical sidebands with frequencies of $f_1 \pm (2f_C - f_B + 2kt)$ are output by the ODPF, which is then amplified by an erbium-doped fiber amplifier (EDFA, Amonics AEDFA-PA-35-B-FA) with a small signal gain of 35 dB. The amplified optical signal is sent to an optical band-pass filter (OBPF, EXFO XTM-50) to suppress the amplified spontaneous emission (ASE) noise. The output optical signal from the OBPF is split into two paths by $OC_4$: One output of $OC_4$ is injected into $MZM_1$ (Fujitsu FTM7938EZ) as an optical reference for radar de-chirping; The other output of the $OC_4$ is connected to high-speed $PD_1$ (u2t MPRV1331A), where the two ±2nd-order optical sidebands beat each other to generate the frequency-quadrupled LFM signal from 24.4 GHz to 28.4 GHz. $EA_2$ (CENTELLAX OA4MVM2) and $EBPF_2$ (Shanghai AT Microwave AT-BPF-2432) are used to boost the signal power and suppress the interference and noise. Then the frequency-quadrupled LFM signal is radiated through a transmitting antenna for contactless respiration and heartbeat monitoring. The echo signal reflected by the human chest is collected through a receiving antenna and then amplified by $EA_3$ (CENTELLAX OA4MVM3) with a gain of 27 dB. The optical reference is modulated by the amplified echo signal at $MZM_1$. After detection in low-speed $PD_4$ (Nortel PP-10G), microwave photonic de-chirping is implemented. The radar de-chirped signal from $PD_4$ and the aforementioned electrical signals from $PD_2$ and $PD_3$ are sampled by a real-time oscilloscope (OSC, Rohde & Schwarz RTO2032). By analyzing the phase of the radar de-chirped signal in Matlab, contactless respiration and heartbeat monitoring for multiple people can be achieved simultaneously.

**3.2 Simultaneously contact and contactless vital sign monitoring using one channel**

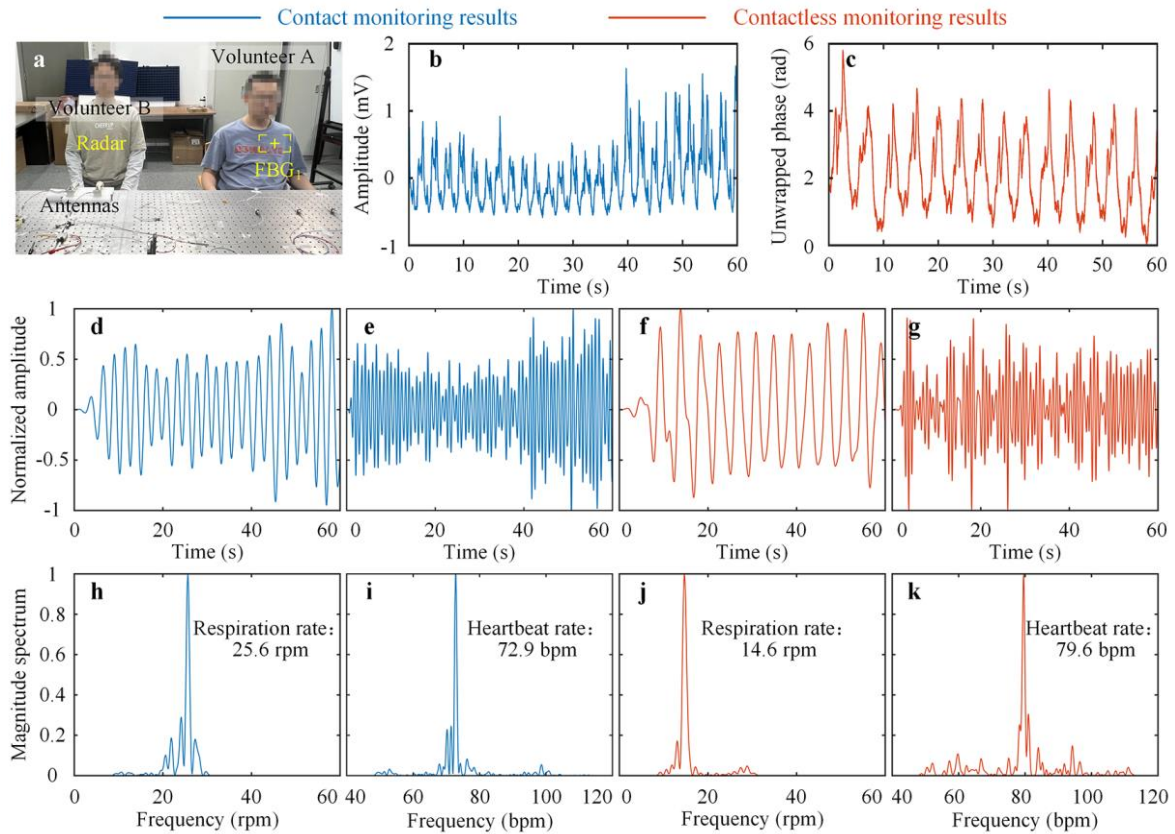

**Fig. 5 Results of simultaneous contact and contactless vital sign monitoring using one channel. a** Two volunteers for contact and contactless vital sign monitoring. **b** FBG-based contact respiration and heartbeat signal from $PD_3$. **c** Radar-based contactless respiration and heartbeat signal from $PD_4$ after using arctangent demodulation algorithm. **d** and **h** Contact respiration waveform and spectrum. **e** and **i** Contact heartbeat waveform and spectrum. **f** and **j** Contactless respiration waveform and spectrum. **g** and **k** Contactless heartbeat waveform and spectrum.

Figure 5a shows a photograph of the respiration and heartbeat monitoring experiment. In this experiment, only the channel corresponding to 1549.36 nm is used for both contact and contactless vital sign monitoring. Two volunteers are invited: One for contact monitoring and the other for contactless monitoring. $FBG_1$ is pasted on the chest of Volunteer A, and Volunteer B is 0.88 m away from the radar transmitting antenna. As the chest wall movements stretch and compress $FBG_1$, the shifts in the Bragg wavelength are translated into variations in the optical intensity of the $FBG_1$ output, which is measured after photodetection in $PD_3$, with the waveform shown in Fig. 5b. In contact vital sign signal processing, the respiration waveform is extracted by a digital bandpass elliptic filter with a passband from 0.13 Hz to 0.5 Hz. The waveform and spectrum of the respiration signal after filtering are shown in Figs. 5d and 5h. Another digital bandpass elliptic filter with a passband from 0.8 Hz to 1.9 Hz is employed to extract the heartbeat signal, whose waveform and spectrum are depicted in Figs. 5e and 5i. It can be observed that the respiration and heartbeat rates of Volunteer A are 25.6 rpm and 72.9 bpm. Compared to the standard monitored respiration rate of 24 rpm and heartbeat rate of 73 bpm, the absolute measurement errors for respiration and heartbeat rates amount to 1.6 rpm and 0.1 bpm, respectively.

In contactless vital sign signal processing, the arctangent demodulation algorithm is utilized to obtain the radar-based respiration and heartbeat signals, as depicted in Fig. 5c. It is obvious that the unwrapped phase changes over time, which is caused by the respiration and heartbeat of the volunteer B. Similar to the contact respiration and heartbeat signal separation method, two digital bandpass elliptic filters are employed to extract the respiration signal and heartbeat signal. The waveform and spectrum of the contactless respiration signal are shown in Figs. 5f and 5j. The respiration rate of Volunteer B is 14.6 rpm, slightly lower than Volunteer A. Figures 5g and 5k depict the waveform and spectrum of the contactless heartbeat signal. The peak of the spectrum corresponds to a heartbeat rate of 79.6 bpm. Compared to the standard monitored respiration rate of 15 rpm and heartbeat rate of 81 bpm, the absolute measurement errors for the respiration and heartbeat rates are 0.4 rpm and 1.4 bpm, respectively. It is worth noting that the aforementioned standard monitored respiration rate is determined by two volunteers who record the number of inhalations and exhalations themselves, while the standard heartbeat rate is determined by monitoring the pulse rate, which is also known as pulse-taking in traditional Chinese medicine.

**3.3 Respiration and heartbeat monitoring of multiple people using two channels**

To further verify the proposed system for the distribution of optical vital sign monitoring signals in a WDM optical network, in this experiment, the two channels corresponding to 1549.36 nm and 1549.92 nm are jointly used: The former for simultaneous contact and contactless vital sign monitoring while the latter for only contact vital sign monitoring. The 1549.92-nm channel is only used for contact vital sign monitoring because the laboratory does not have additional optical filters to build the subsequent radar detection part of this channel. Under these circumstances, the respiration and heartbeat monitoring experiment of multiple people is implemented. Since the OSC only has two channels, it cannot simultaneously capture the two outputs of $PD_2$ and $PD_3$ and the radar de-chirped signal from $PD_4$. Therefore, in the experiment, the data acquisition of the three PDs is divided into two steps: The first step is that the optical intensity of $FBG_2$ from $PD_2$ and the radar de-chirped signal from $PD_4$ are sampled to monitor the respiration and heartbeat for three people; The second step is that optical intensities of $FBG_1$ and $FBG_2$ from $PD_3$ and $PD_2$ are sampled to achieve contact respiration and heartbeat monitoring for two people.

First, Volunteer A serves as the contact respiration and heartbeat monitoring subject with $FBG_2$ placed on his chest while Volunteers B and C serve as the contactless respiration and heartbeat monitoring subjects. Along the radar line of sight, Volunteers B and C sit on stools 1 m and 1.65 m away from the antenna, respectively. The specific positions of the three volunteers are shown in Fig. 6a. In contact respiration and heartbeat monitoring for Volunteer A, one bandpass elliptic filter with a passband from 0.13 Hz to 0.5 Hz is used to extract the respiration signal, and another bandpass elliptic filter with a passband from 0.8 Hz to 1.9 Hz is utilized to acquire the heartbeat signal. The respiration waveform and spectrum are shown in Figs. 6c and 6f. It can be seen that there is an obvious peak in the spectrum of the respiration signal, and the respiration rate corresponding to the peak is 20.8 rpm. Figures 6i and 6l show the heartbeat waveform and spectrum. His heartbeat rate is 86.5 bpm, which is very close to the actual value of 87 bpm.

In contactless respiration and heartbeat monitoring for Volunteers B and C, the radar de-chirped signal is further processed. Figure 6b shows the range profile for the two volunteers, obtained by performing Fast Fourier Transforms (FFT) and peak search on the radar de-chirped signal. The two peaks are easily observed, and the distances corresponding to the peaks are 1.01 m and 1.68 m, which are very close to the set values. The difference is caused by the respiration and heartbeat, as well as the radar range resolution. Two digital filters are used to select each of the range peaks and the arctangent demodulation algorithm is employed to extract the phases corresponding to one volunteer. Phase unwrapping is implemented to acquire continuous phase changes. The respiration and heartbeat signals are then obtained through two digital bandpass elliptic filters. Figures 6d and 6g depict the respiration waveform and spectrum of Volunteer B, while Figures 6j and 6m show the heartbeat waveform and spectrum of Volunteer B. As can be seen, there is a second respiration harmonic in Volunteer B, which does not interfere with heartbeat rate monitoring because the harmonic frequency is much lower than the fundamental frequency. The respiration waveform and spectrum of Volunteer C are shown in Figs. 6e and 6h. A clear peak is observed, which represents the respiration rate of Volunteer C. Figures 6k and 6n show the heartbeat waveform and spectrum. The noise floor of Volunteer C's heartbeat signal is noticeably higher than that of Volunteer B's, primarily because Volunteer C's position, which is situated further away from the radar antenna, leads to a weaker echo. The monitored and actual values of the respiration and heartbeat rates of the three volunteers are shown in **Table 1**. The maximum absolute measurement errors of respiration and heartbeat rates are 0.6 rpm and 1.2 bpm, respectively.

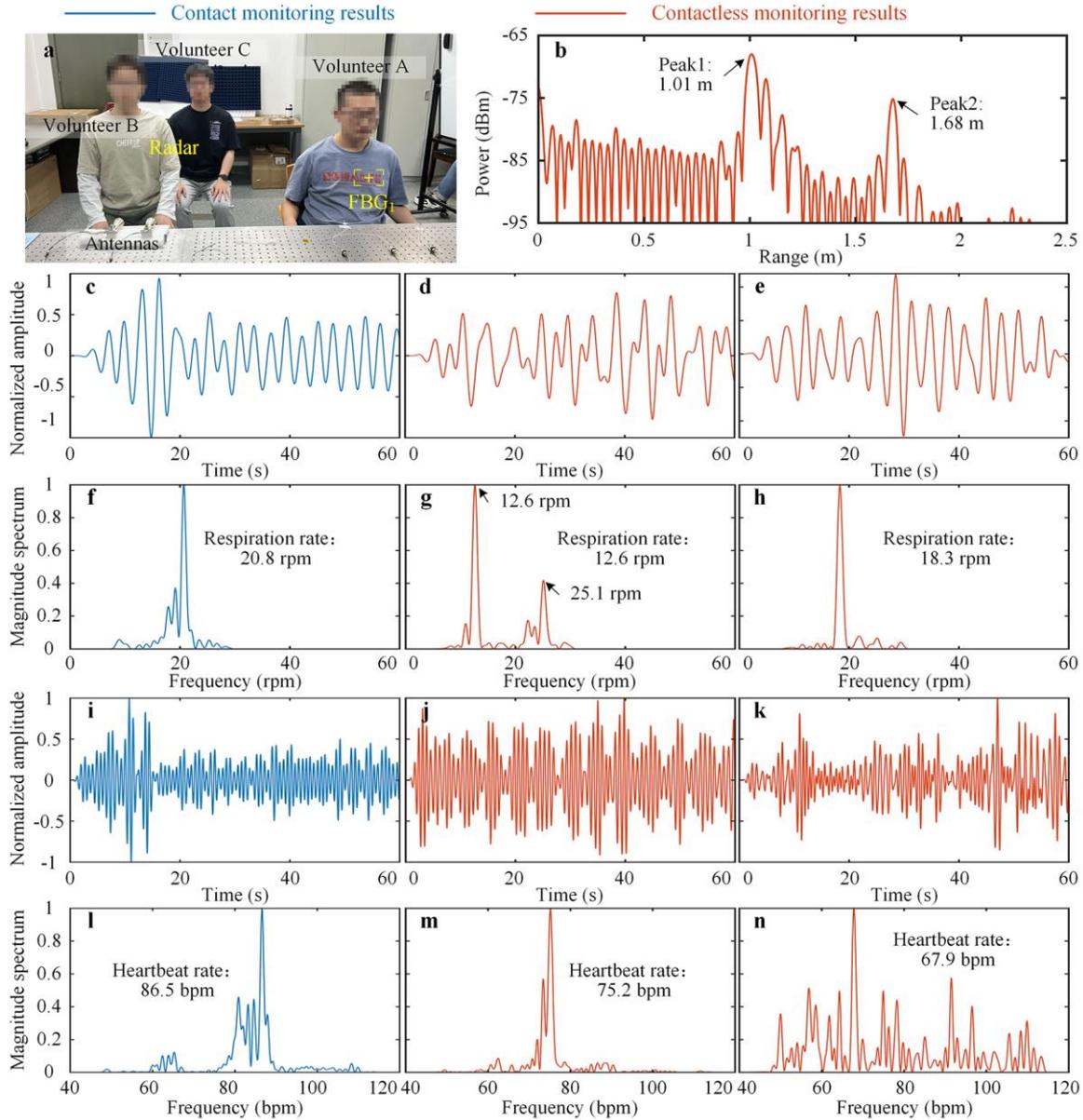

**Fig. 6 Results of simultaneous contact and contactless vital sign monitoring using two channels. a** Three volunteers for contact and contactless vital sign monitoring. **b** Range profile of Volunteers B and C obtained from the radar de-chirped signal. **c** and **f** Contact respiration waveform and spectrum of Volunteer A. **i** and **l** Contact heartbeat waveform and spectrum of Volunteer A. **d** and **g** Contactless respiration waveform and spectrum of Volunteer B. **j** and **m** Contactless heartbeat waveform and spectrum of Volunteer B. **e** and **h** Contactless respiration waveform and spectrum of Volunteer C. **k** and **n** Contactless heartbeat waveform and spectrum of Volunteer C.

**Table 1 Comparison of monitored and actual values of respiration and heartbeat rates**

| | Age (years) | Respiration rate [rpm] | | | Heartbeat rate [bpm] | | |
|---|---|---|---|---|---|---|---|
| | | Monitored value | Actual value | Error | Monitored value | Actual value | Error |
| Volunteer A | 28 | 20.8 | 21 | -0.2 | 86.5 | 87 | -0.5 |
| Volunteer B | 27 | 12.6 | 12 | 0.6 | 75.2 | 74 | 1.2 |
| Volunteer C | 27 | 18.3 | 18 | 0.3 | 67.9 | 68 | 0.1 |

Finally, simultaneously contact vital sign monitoring for two people using both the two channels corresponding to 1549.36 nm and 1549.92 nm is demonstrated. $FBG_1$ and $FBG_2$ are pasted on the chests of Volunteers A and B, respectively. The optical intensities of $FBG_1$ and $FBG_2$ output are sampled five times. Two digital bandpass elliptic filters are utilized to extract the respiration and heartbeat signals of the two volunteers. Figures 7a and 7d show the respiration and heartbeat waveforms of the first sample of Volunteer A. The respiration and heartbeat waveforms of the first sample of Volunteer B are shown in Figs. 7b and 7e. Respiration and heartbeat rates are extracted by performing FFTs and peak searches on the respiration and heartbeat waveforms. Figures 7c and 7f depict the monitored values and actual values of the two volunteers. The monitored values of respiration and heartbeat rates accord with the actual values and the maximum absolute measurement errors of respiration and heartbeat rates are 1.4 rpm and 2.3 bpm, respectively.

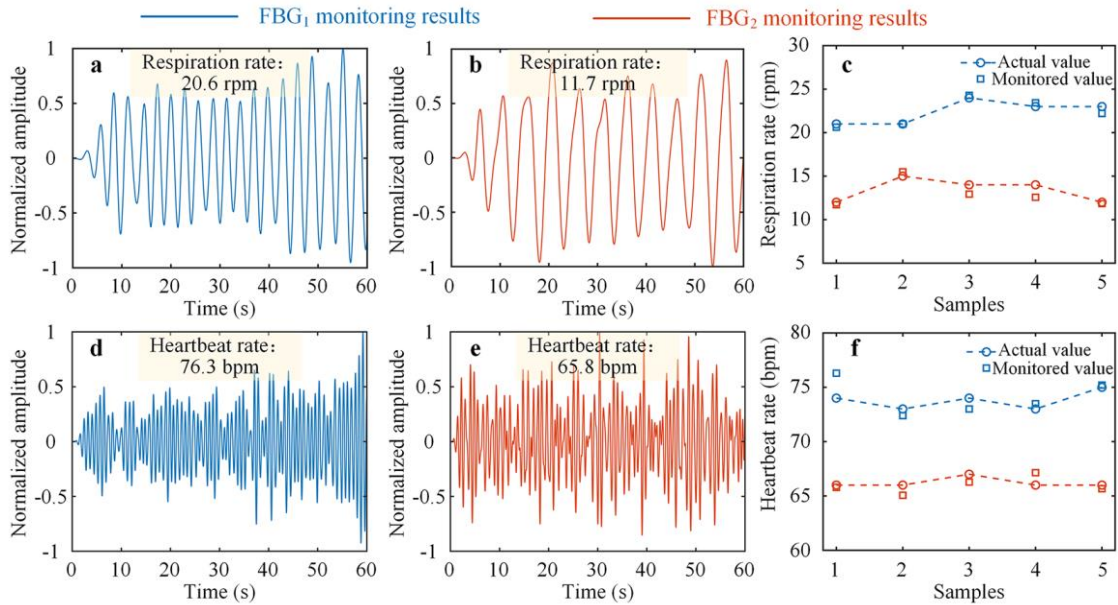

**Fig. 7 Results of simultaneous contact vital sign monitoring for two people. a** Respiration and **d** heartbeat waveforms of the first sample of Volunteer A. **b** Respiration and **e** heartbeat waveforms of the first sample of Volunteer B. Monitored and actual values of respiration and heartbeat rates for **c** Volunteer A and **f** Volunteer B.

## 4. Discussion

In the above respiration and heartbeat monitoring experiments, the optical intensity of the FBG output and the radar de-chirped signal are sampled with the same sampling time of 60 s, which means that users need to spend up to 60 s for monitoring. In many application scenarios, it is highly desirable that the monitoring can be accomplished in a much faster manner. Therefore, a deeper analysis of the data is conducted to study the influence of sampling time on the respiration and heartbeat monitoring results. The two sets of data respectively corresponding to the contact monitoring and contactless monitoring in Fig. 5b and 5c are used. By intercepting sampling signals of different time lengths, including 5 s, 10 s, 20 s, 30 s, 40 s, 50 s, and 60 s, we further analyze the monitoring results under different sampling time lengths. The signal processing method is consistent with that described above. In this study, the 3-dB bandwidth of the respiration and heartbeat signal spectrum is used to indicate the respiration resolution and heartbeat resolution. Figures 8a and 8d shows contact respiration resolution and heartbeat resolution, while Figs. 8b and 8e shows contactless respiration resolution and heartbeat resolution. It is observed that with the increase of sampling time length, respiration resolution and heartbeat resolution are obviously improved. When the sampling time length exceeds 30 s, the respiration resolution and heartbeat resolution tend to stabilize. Figures 8c and 8f depict the monitored and actual values of the respiration and heartbeat rates at different sampling time lengths. Although the monitoring resolution decreases with the reduction of sampling time length, the decline in monitoring resolution does not have a significant impact on the monitoring results since there is often only one peak in the monitoring of respiration and heartbeat. According to the above analysis and results, the sampling time length of the proposed respiration and heartbeat monitoring system can be further shortened to even 5 s to quickly complete the monitoring of respiration and heartbeat, thereby improving the efficiency and real-time performance of the system.

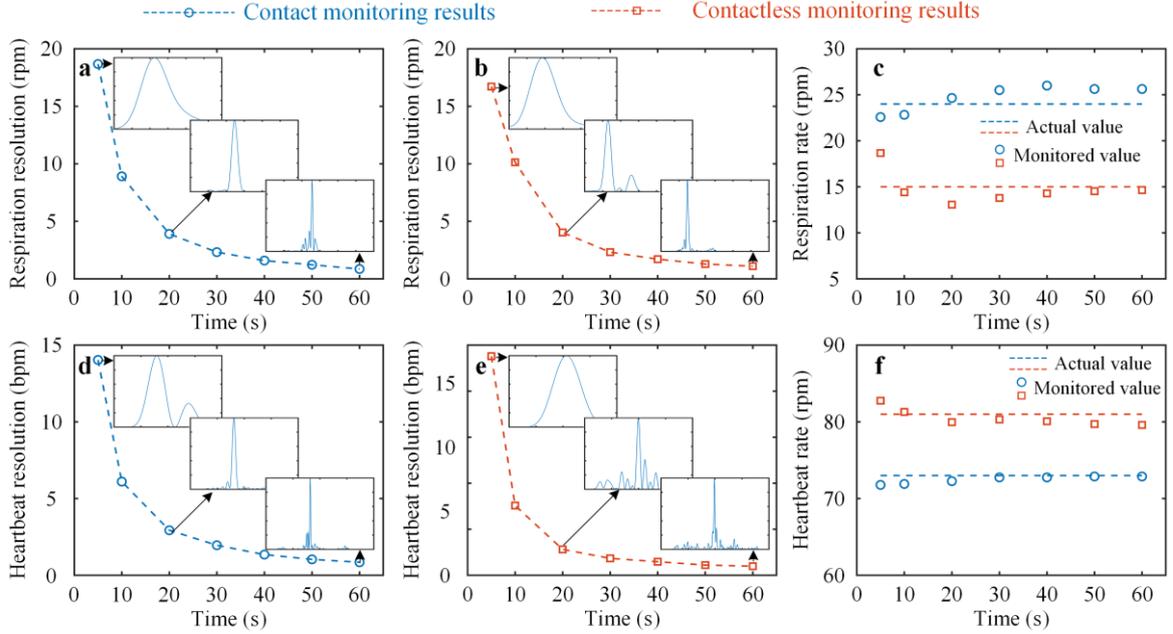

**Fig. 8 Results of contact and contactless vital sign monitoring at different sampling time lengths.** Contact **a** respiration resolution and **d** heartbeat resolution. Contactless **b** respiration resolution and **e** heartbeat resolution. **c** contact and **f** contactless monitored and actual values of the respiration and heartbeat rates. The horizontal axis of the inset (**a**, **b**, **d**, and **e**) diagram represents frequency, whereas the vertical axis represents the normalized magnitude spectrum.

As shown in Fig. 4, the above experiments only simulate the application scenario of deploying two monitoring channels at the user terminal through two wavelengths. Theoretically, one channel (i.e., a single wavelength) can achieve contact monitoring of one subject using the FBG sensor and simultaneous contactless monitoring of multiple subjects using the radar sensor, as shown in the results presented in Fig. 5. By simultaneously deploying two wavelengths, the system is capable of implementing contact monitoring of two subjects and simultaneous contactless monitoring of multiple subjects, as shown in the results presented in Figs. 6 and 7. It should be noted that the second channel in this article is only used for contact monitoring and is not equipped with contactless monitoring functionality at the same time. This is partially due to the limitations in experimental conditions and partially because radar-based contactless monitoring itself has the ability to monitor multiple subjects simultaneously through a single channel. In practical applications, it is generally sufficient to equip a user terminal with one WDM wavelength, which enables both contact and contactless monitoring functions. For some application scenarios that require contact monitoring of multiple subjects simultaneously, multiple WDM wavelengths can be equipped for them. In most cases, even if multiple wavelengths are equipped, it is sufficient to configure contactless monitoring capabilities for only one of the wavelengths. Of course, in special application scenarios, it is also possible to configure contactless monitoring capabilities for multiple wavelengths to achieve multiple radar transceiver architectures. However, the cost of the system will be relatively high at this time. The user terminal of the system proposed in this article is directly connected to the WDM optical network through optical fibers, and the number of wavelengths can be requested based on demand, making it highly scalable and compatible with existing WDM optical networks.

In summary, we have proposed and experimentally demonstrated a seamlessly integrated contact and contactless vital sign monitoring system with ROF assistance, which can simultaneously implement respiration and heartbeat monitoring for multiple people. To the best of our knowledge, this is the first photonic demonstration for simultaneous contact and contactless vital sign monitoring, which also features its perfect integration with the existing sophisticated WDM optical network infrastructure and ROF technique, and offers significant advantages in terms of cost-effectiveness and scalability for large-scale deployment. A proof-of-concept experiment is performed. Contact and contactless respiration and heartbeat monitoring of three people is simultaneously realized. During a monitoring time of 60 s, the maximum absolute measurement errors of respiration and heartbeat rates are 1.6 rpm and 2.3 bpm, respectively. The measurement error does not have an obvious change even when the monitoring time is decreased to 5 s. The research provides a potential solution to design a seamlessly integrated contact and contactless vital sign monitoring system for large hospitals, nursing homes, and smart communities. This research holds vast application prospects in future smart and healthy city scenarios, meeting the diverse health inspection and monitoring needs of future hospitals, nursing homes, and smart communities. At the same time, due to its high degree of integration with the WDM optical network, it is simple to deploy and easy to expand. In addition, due to the adoption of a centralized deployment for complex equipment, the cost of deploying each distributed user terminal is very low.

## 5. Materials and methods

### 5.1 Transmission responses of the two FBGs

In the system, the strain-sensing characteristic of FBGs is utilized to monitor one's respiration and heartbeat rates. To prevent the FBG from being damaged, $FBG_1$ and $FBG_2$ are all encapsulated with Scotch tape, with one of them before being encapsulated shown in Fig. 9a. The transmission spectra of the two FBGs are measured using an ASE source (Max-Ray EDFA-PA-35-B) and an optical spectrum analyzer (OSA, ANDO AQ6317B). Figures 9b and 9c depict the transmission spectra of $FBG_1$ and $FBG_2$. The 3-dB bandwidths of the two FBGs are 11.2 GHz and 10 GHz, and the two notches are 17.70 dB and 17.76 dB lower than the passband, respectively.

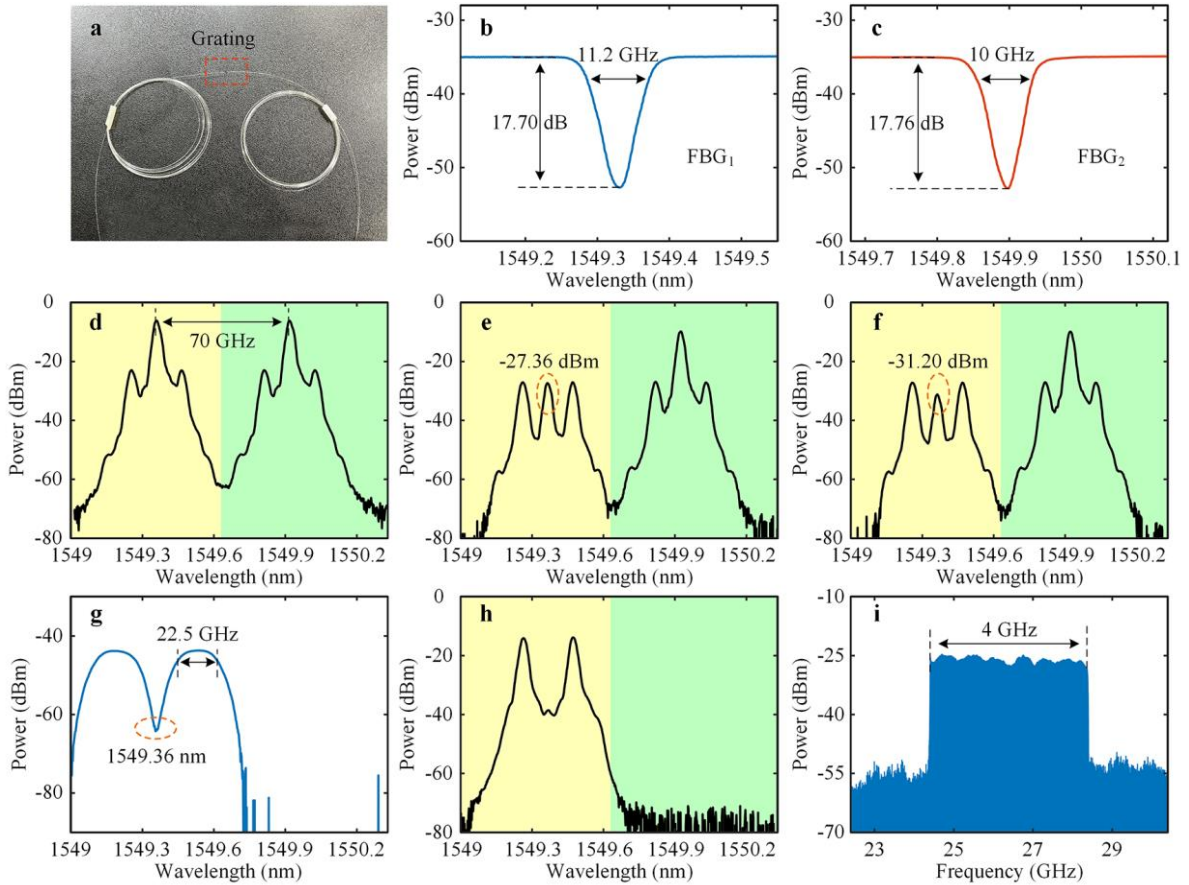

**Fig. 9 Optical spectra and transmitted radar signal. a** FBG sensor. Transmission spectra of **b** $FBG_1$ and **c** $FBG_2$. **d** Optical spectrum of the IF-LFM-modulated signal from $MZM_0$. **e** and **f** Optical spectra of $FBG_1$ output signal when the Bragg wavelength shifts with the movement of the human chest wall. **g** Frequency response of the ODBF. **h** Optical spectrum of the input signal of $PD_1$. **i** Electrical spectrum of the generated frequency-quadrupled radar signal.

### 5.2 Broadband radar signal generation

The microwave up-conversion module is constructed to generate an IF-LFM signal. First, the AWG is used to generate a low-frequency LFM signal with a center frequency of 1.75 GHz, a bandwidth of 1 GHz, a pulse period of 100 $\mu$s, and a pulse width of 60 $\mu$s. The low-frequency LFM signal is mixed with a 13.5-dBm and 8.35-GHz local oscillator (LO) signal from the MSG, and then amplified by $EA_1$ and filtered by $EBPF_1$. Thus an IF-LFM signal with a 6.6-GHz center frequency and a 1-GHz bandwidth is generated and applied to $MZM_0$. The optical spectrum of the IF-LFM-modulated optical signal from $MZM_0$ is shown in Fig. 9d. The frequency difference between the two optical carriers is 70 GHz. The Bragg wavelength of $FBG_1$ will shift when the grating period and effective refractive index of $FBG_1$ are changed due to the chest wall movement. By positioning the optical wavelength, centered at 1549.36 nm, at the edge of the transmission spectrum of $FBG_1$, any chest wall movement will correspondingly trigger a change in the transmission power of $FBG_1$. Figures 9e and 9f give the optical spectra of the $FBG_1$ output signal when the Bragg wavelength shifts with the movement of the human chest wall. It is evident that the intensity of the optical carrier, centered at 1549.36 nm, undergoes changes, whereas other components remain consistent and unchanged.

An ODPF, with its frequency response shown in Fig. 9g, is used to select ±2nd-order optical sidebands of the IF-LFM signal. As can be seen, the 3-dB bandwidth of the ODPF is 22.5 GHz and the notch position is consistent with the 1549.36-nm wavelength. The output optical signal from the ODPF is amplified by the EDFA and filtered by the OBPF as the optical reference and the input signal of $PD_1$. The optical spectrum of the input signal of $PD_1$ is shown in Fig. 9h. Only two ±2nd-order optical sidebands are observed, while the optical carrier in this channel and the optical signal in the other channel are effectively suppressed. After the optical signal shown in Fig. 9h is detected in $PD_1$, amplified in $EA_2$, and filtered in $EBPF_2$, the electrical spectrum of the generated radar signal is shown in Fig. 9i. In contrast to the IF-LFM signal, the transmitted radar signal exhibits a fourfold enhancement in both its center frequency and bandwidth, boasting a center frequency of 26.4 GHz and a bandwidth of 4 GHz.

### 5.3 Contact and contactless signal sampling

The proposed vital sign monitoring system can simultaneously realize contact and contactless respiration and heartbeat monitoring. In the experiment, a microwave up-conversion module is utilized to generate the IF-LFM signal with a center frequency of 6.6 GHz, a bandwidth of 1 GHz, a pulse period of 100 $\mu$s, and a pulse width of 60 $\mu$s. A 10-MHz sinusoidal signal from MSG is sent to the AWG for signal synchronization. In the process of FBG-based contact respiration and heartbeat monitoring, the movement of one's chest wall resulting from respiration and heartbeat directly correlates to variations in the optical intensity of the FBG output, which is converted to the electrical domain via $PD_2$ and $PD_3$. In the experiment, outputs of $PD_2$ and $PD_3$ are sampled by the OSC under a sampling rate of 50 Sa/s. The waveforms are then processed in the digital domain to extract the respiration and heartbeat rates. In the process of radar-based contactless respiration and heartbeat monitoring, the generated broadband LFM signal from 24.4 GHz to 28.4 GHz is used as the transmitted radar signal. The radar de-chirped signal is obtained after photonic frequency mixing in $PD_4$ and sampled using the OSC under a sampling rate of 10 MSa/s. By analyzing the phase of the radar de-chirped signal, the respiration and heartbeat rates in contactless monitoring can be obtained. It should be noted that for contactless monitoring, since the radar de-chirped signal is being sampled and the sampling rate is high, direct sampling would result in issues such as excessively large amounts of sampled data and slow processing speeds. Since the changes in respiration and heartbeat rates are slow, it is not necessary to sample all radar de-chirped waveforms throughout the entire sampling time length. This can significantly reduce the amount of sampled data and increase processing speed. To do this, a rectangular pulse signal with a pulse period of 20 ms and a pulse width of 60 $\mu$s is connected to the external trigger port of the OSC to implement intermittent sampling of the signal. In this experiment, the rectangular pulse is directly generated from the AWG. It should be noted that, in practical applications, the rectangular pulse can be generated locally in the user terminal by a pulse source.

### Acknowledgements

This work was supported by the National Natural Science Foundation of China under Grant 62371191 and Grant 61971193.